\documentclass[12pt]{article}

\usepackage{setspace}
\usepackage{graphics}
\usepackage{latexsym}
\usepackage{rotating}
\usepackage{amsmath}
\usepackage{color}
\usepackage{mathpazo}
\usepackage{fullpage}
\usepackage{psfrag}
\usepackage{pifont}
\usepackage{amssymb}
\usepackage{indentfirst}
\usepackage{slidesec}
\usepackage{graphicx,subfigure}
\usepackage{epsfig}
\usepackage{fancybox}
\usepackage{color}
\usepackage{amssymb,amsmath}
\usepackage{epsfig, graphicx}
\usepackage{epsfig}
\usepackage{mathrsfs}
\usepackage{verbatim}
\usepackage{xcolor}
\usepackage{xspace}

\usepackage{bbm}
\usepackage{natbib}
\bibpunct[, ]{(}{)}{,}{a}{}{;}
\usepackage{ctable}
\usepackage{footnote}
\usepackage{enumerate}
\makesavenoteenv{tabular}

\newtheorem{remark}{Remark}

\def \ba {\textbf{a}}
\def \bD{\textbf{D}}
\def \bZ {\textbf{Z}}
\def \bM {\textbf{M}}
\def \bV {\textbf{V}}
\def \bR {\textbf{R}}

\def \bbeta{\boldsymbol{\beta}}
\def \bgamma{\boldsymbol{\gamma}}

\def\alphahat{\widehat{\alpha}}
\def\rhohat{\widehat{\rho}}
\def\thetahat{\widehat{\theta}}
\def \phat {\widehat{p}}
\def \what {\widehat{w}}
\def \nfrak{\mathfrak{n}}
\def \Dfrak{\mathfrak{D}}
\def \Ffrak{\mathfrak{F}}
\def \bbetahat{\widehat{\bbeta}}

\def \Gsc{\mathcal{G}}
\def \Wsc{\mathcal{W}}
\def \Rsc{\mathcal{R}}
\def \Vsc{\mathcal{V}}
\def \Gschat{\widehat{\Gsc}}
\def \Wschat{\widehat{\Wsc}}

\newcommand{\auc}{\mbox{AUC}}
\newcommand{\Cov}{\mbox{Cov}}

\newcommand{\auchat}{\widehat{\auc}}

\def \Tdag{T^{\dag}}

\def\Rscr{\mathscr{R}}

\def \sumiN{\sum_{i=1}^{N}}
\def \sumjN{\sum_{j=1}^{N}}

\def \Pbbm{\mathbbm{P}}
\def\Pbbmhat{\widehat{\Pbbm}}

\def\transpose{{\sf \scriptscriptstyle{T}}}
\def\trans{^{\transpose}}
\def \whatast{\what^{\ast}}
\def\omegahat{\widehat{\omega}}
\def\omegahatast{\omegahat^\ast}

\def\sumiN{\sum_{i=1}^N}
\def\sumjN{\sum_{j=1}^N}
\def\Gschatast{\Gschat^\ast}

\def \E{\text{E}}
\def \Var{\text{Var}}
\def \sauc{\mbox{\tiny $\auc$}}
\def \sgamma{\mbox{\tiny $\bgamma$}}
\def \stheta{\mbox{\tiny $\theta$}}
\def \bCsc{\boldsymbol{\mathcal C}}
\def \wbre{\breve{w}}
\def \thetabre{\breve{\theta}}
\def \thetahat{\widehat{\theta}}
\def \tpr{\text{TPR}}
\def \fpr{\text{FPR}}
\def \npv{\text{NPV}}
\def \ppv{\text{PPV}}
\def \auc {\text{AUC}}
\def \tprhat{\widehat{\tpr}}
\def \fprhat{\widehat{\fpr}}
\def \npvhat{\widehat{\npv}}
\def \ppvhat{\widehat{\ppv}}
\def \auchat {\widehat{\auc}}

\begin{document}

\title{A new weighting method when not all the events are selected as cases in a nested case-control study}
\date{}
\author{Qian M. Zhou$^{a}$, Xuan Wang$^{b}$, Yingye Zheng$^{c}$, and Tianxi Cai$^{d}$ \vspace{0.5cm} \\
\small $^{a}$ Department of Mathematics and Statistics, Mississippi State University,\\
\small  Starkville, Mississippi 39762, USA \\
\small $^{b}$ School of Mathematical Sciences, Zhejiang University,\\
\small Hangzhou, Zhejiang, 310027, China \\
\small $^{c}$ Public Health Sciences Division, Fred Hutchinson Research Center,\\
\small Seattle, WA, 98109, USA\\
\small $^{d}$ Harvard T.H. Chan School of Public Health,\\
\small Boston, MA, 02115, USA
}

\maketitle

\begin{abstract}
Nested case-control (NCC) is a sampling method widely used for developing and evaluating risk models with expensive biomarkers on large prospective cohort studies. The biomarker values are typically obtained on a sub-cohort, consisting of all the events and a subset of non-events. However, when the number of events is not small, it might not be affordable to measure the biomarkers on all of them. Due to the costs and limited availability of bio-specimens, only a subset of events is selected to the sub-cohort as cases. For these ``untypical" NCC studies, we propose a new weighting method for the inverse probability weighted (IPW) estimation. We also design a perturbation method to estimate the variance of the IPW estimator with our new weights. It accounts for between-subject correlations induced by the sampling processes for both cases and controls through perturbing their sampling indicator variables, and thus, captures all the variations. Furthermore, we demonstrate, analytically and numerically, that when cases consist of only a subset of events, our new weight produces more efficient IPW estimators than the weight proposed in \cite{samuelsen1997psudolikelihood} for a standard NCC design. We illustrate the estimating procedure with a study that aims to evaluate a biomarker-based risk prediction model using the Framingham cohort study.  

\end{abstract}


\section{Introduction}

Risk prediction using novel biomarkers plays a vital role in disease prevention and disease management. The development and evaluation of risk models require rich information from large-scale cohort studies, where the participants are followed prospectively to the clinical outcome of interest, and their clinical information is collected at baseline. Many cohorts also obtain biological specimens which are used later for investigating new biomarkers to improve the predictive capacity of the risk model. Due to the cost and effort for ascertaining biomarkers and the need to preserve precious biologic samples, a two-phase study is often conducted in this setting,  where new biomarkers are typically measured on a sub-cohort, instead of the entire cohort. 

Careful planning and analysis are needed for inferring the results based on such a sub-cohort sampling to avoid biased conclusions and improve study efficiency.  For example, under a classic nested case-control (NCC) design, all patients who have encountered the event (cases) will be selected, and for each case, a number of matched controls will be selected, among those who are event-free at the failure time of the case, i.e., from the so-called risk set of the case. Conditional logistic regression \citep{goldstein1992asymptotic} has been used to estimate hazard ratios under the Cox proportional hazards (PH) model \citep{cox1972regression}, but  the resulting estimators are inefficient. Additionally, this method cannot be applied to other models. \cite{scheike2004maximum} and \cite{zeng2006efficient} proposed a maximum likelihood estimator (MLE) that attains the semi-parametric efficiency bound. However, it requires the estimation of the conditional density of the new biomarkers given other observed variables and an assumption of censoring being independent of the markers. The  inverse probability weighted (IPW) estimators have been considered for estimating the model parameters \citep{samuelsen1997psudolikelihood} and accuracy summaries of a risk prediction model \citep{cai2012evaluating, cai2011nonparametric, zhou2015assessing}. With an IPW estimator, the weights for cases are fixed at 1, and the weight for a selected control is the inverse of the probability that the participant is selected to the sub-cohort. For IPW estimators under NCC designs, obtaining an analytical variance estimation is challenging because the variance expression is complicated, involving calculating covariances among sampled individuals. Therefore, the IPW methods have been limited to NCC designs where all the events are cases in the sub-cohort.  

In this manuscript, we consider developing and validating a risk prediction model with a variation of the NCC sampling, where only a subset of events are sampled as cases, and controls are selected from the risk sets of the selected cases. This setting can occur when the number of events is not small, and there is a practical constraint for using all the events to save cost or samples.
For example, \cite{jakszyn2006endogenous} measured antibodies to Helicobacter pylori and vitamin C levels on 229 out of 314 gastric cancer patients from the European Prospective Investigation into Cancer and Nutrition study. For each case, two to four controls were selected; the same design was employed in \cite{jakszyn2012dietary}. \cite{lu2018multi} conducted a multi-center study on hospitalization costs and length of stay due to healthcare-associated infection (HAI) using the HAI prevalence survey in Sichuan in 2016. They selected 10 cases from each of the 51 hospitals included in the study. For each case, one control is selected and matched on several criteria. 

The design used in the above examples can be considered as an ``untypical” NCC design: the sub-cohort cases consist of only a subset of participants with events. With such a design, to estimate the risk model parameters and accuracy measures for model evaluation, one can not directly apply the existing estimation methods that were proposed for the standard NCC studies where all the events are selected as cases. To this end, we propose a new weight, which leads to a more efficient and robust estimation, compared to the weights used for the standard NCC design as originally proposed by \cite{samuelsen1997psudolikelihood}. We demonstrate, analytically and numerically, that when only a subset of events are cases, our proposed weight leads to improved performance compared to Samuelsen's weight. When all the events are cases, these two weights are equivalent. Thus, our method provides a powerful tool for analyzing all types of NCC designs, no matter the cases comprise all the events or just a subset.
Compared to the conditional logistic regression and MLE-based estimators, the IPW estimators can be easily extended beyond the Cox model. In this article, we illustrate our method under both the Cox model and the time-dependent generalized linear model (GLM) \citep{uno2007evaluating}.  In addition, we extend the procedures to  estimate accuracy measures, such as the true positive rate (TPR), the false positive rate (FPR), the area under a receiver operating characteristic (ROC) curve (AUC), the positive predictive value (PPV), and the negative predictive value (NPV) \citep{heagerty2005survival, cai2012evaluating,cai2011nonparametric,zhou2015assessing}.

Furthermore, we conduct a rigorous theoretical study of the proposed weighted estimators and provide a valid variance estimator. Standard resampling procedures such as bootstrap fail to capture the correlation structure induced by the finite-population sampling \citep{gray2009weighted}. \cite{cai2013resampling} proposed a perturbation resampling method for NCC data in which all the events are cases. This method approximates the between-control correlations via perturbing their sampling indicator variables. However, when only a subset of events are selected as cases, their method cannot account for the variances and correlations caused by the sampling process for cases. To address this issue, we propose a new perturbation procedure that is generally applicable for variations under the NCC sampling scheme. 

The remaining manuscript is organized as follows. In Section 2, we introduce the new weight. We outline the general structure of the asymptotic variance for the IPW estimator with the new weight. Additionally, we present the perturbation resampling procedure for the variance estimation. Our new weight is compared with Samuelsen's weight analytically in Section 3 and numerically in Section 4. Concluding remarks are given in Section \ref{sec:discussion}.

\section{Model Specification and Estimation}\label{sec:IPWweights}

\subsection{Notation}
Let $\Tdag_i$ denote the time to the event of interest, $i=1,\cdots,N$. Due to censoring, we only observe $T_i=\min(\Tdag_i,C_i)$ and $\delta_i = I(\Tdag_i \leq C_i)$, where $C_i$ is the censoring time, and $I(\cdot)$ is an indicator function. The subjects with $\delta_i=1$ are referred to as events, and those with $\delta_i=0$ are referred to as non-events. Let $\bZ_i$ denote a $p$-dimensional vector of markers, including the clinical markers and biomarkers. The relationship between $T$ and $\bZ$ can be specified with a regression model with details given in Section \ref{model}, 

We also define the following NCC sampling indicator variables. Let $V_{1j}=1$ if subject $j$ is selected to the sub-cohort as a \textit{case}. Let $\pi_1$ denote the proportion of events that are cases. When $\pi_1=1$, all the events are cases, and $V_{1j}=1$ if $\delta_j=1$. We define an indicator variable $V_{0j}^i=1$ if subject $j$ is sampled as a \textit{control} of subject $i$, a case of the sub-cohort. Let $V_{0j}=1$ if subject $j$ is selected to the sub-cohort as a control, and it can be expressed as $V_{0j}=1-\prod_{i:j\in \Rscr_{i}}(1-V_{1i}V_{0j}^i)$. Here, $\Rscr_i = \{k: 1\le k \le n, T_k \ge T_i\}$ is the risk set of subject $i$, including all the subjects who have not experienced the event by subject $i$'s event time. For some studies, the controls are also matched to each case on some variables. For these situations, the risk set is expressed as $\Rscr_i = \{k: 1\le k \le n, T_k \ge T_i,|\bM_k-\bM_i| \le \ba_0\}$, where $\bM_i$ is a vector of matching variables, and $|\ba| \le \ba_0$ denotes $|\ba|$ being less than or equal to $\ba_0$ component-wise.

Finally, let $V_j = \delta_iV_{1j} + (1-\delta_iV_{i1})V_{0j}$ indicates whether subject $j$ is ever selected to the sub-cohort.   The values of $\bZ_i$ are  fully observed only if $V_j=1$.

\subsection{The new weight for IPW estimators}\label{sec:weights}
Following the IPW framework for NCC study designs, for the setting where cases are sampled from all participants with $\delta_i =1$, we propose the following weight,
\begin{equation}\label{equ:what}
\what_j = \delta_{j}V_{1j}/\pi_1 + (1-\delta_{j})V_{0j}/\phat_{0j},
\end{equation}
where 
\begin{equation}\label{equ:p0j}
\phat_{0j}=1-\prod_{i:j\in\Rscr_{i}} \left\{1-\frac{m}{\nfrak_{i}-1}\delta_{i}V_{1i}\right\}  
\end{equation}
is the probability of subject $j$ being selected as a control \citep{samuelsen1997psudolikelihood}, $m$ is the number of controls for each case, and $\nfrak_i$ is the size of the risk set $\Rscr_i$ for subject $i$. 

As mentioned earlier, when not all the events are selected as cases, a control could be a non-event or an event. Table \ref{tab:weight-comparison} lists the weight $\what_j$ in equation (\ref{equ:what}) assigned to the following three groups of subjects in the sub-cohort: (i) events that are cases, referred to as \textit{event cases}, (ii) events that are controls, referred to as \textit{event controls}, (iii) non-events that are controls, referred to as \textit{non-event controls}.

\begin{remark}
The probability $\phat_{0j}$ is usually small for large cohort studies because $m$ is often very small relative to the size of the risk set. In addition, when not all the events are cases, the probability $\phat_{0j}$ for an event, say subject $j$, with a short event time can be close to zero, since there are very few cases that subject $j$ is eligible to be included in their risk sets.
\end{remark}

\subsection{IPW estimation}
\label{model}
We consider (i) the Cox PH model, and (ii) the time-dependent GLM model as the risk model. Under each model, we present the IPW estimators for the model parameters and accuracy parameters.

\subsubsection{Model parameters estimation}
Both these two models can be expressed in the following form:
\begin{equation}\label{equ:risk-model}
P(\Tdag_i \le t_0 \mid \bZ_i) = g\left(\alpha_{t_0} + \bbeta_{t_0} \trans\bZ_i\right) \triangleq \Pbbm_{t_0,i} ,
\end{equation}
where $g(\cdot)$ is a link function, and $\bbeta_{t_0}$, a $p$-dimensional vector, are the effects of the markers $\bZ_i$ on the risk $\Pbbm_{t_0,i}$. Let $\bgamma_{t_0}=\left(\alpha_{t_0},\bbeta_{t_0} \trans\right)\trans$.

\paragraph{Cox model.} The Cox PH model can be expressed as 
$$
P(\Tdag_i \le t_0 \mid \bZ_i) = 1-\exp\left[-\exp\left\{ \log \Lambda_0(t_0) +\bbeta \trans\bZ_i \right\}\right],
$$
where $\Lambda_0(t_0)$ is the baseline cumulative hazard function. Based on equation (\ref{equ:risk-model}), $\alpha_{t_0}=\log \Lambda_0(t_0)$, and it can be be estimated by the IPW Breslow’s estimator \citep{cai2012evaluating}  using the proposed weight $\what_j$ in equation (\ref{equ:what}). The marker effects $\bbeta$ are estimated via maximizing the IPW log partial likelihood function \citep{samuelsen1997psudolikelihood} with the weight $\what_j$. 

The Cox PH model assumes the marker effects $\bbeta$ to be constant over time. However, in practice, the biomarkers may have strong effects on the short term risk but weak for the long term risk, or vice versa \citep{zhou2015assessing}.  In these situations, time-dependent GLMs are able to vary the marker effects over time $t_0$. 

\paragraph{Time-dependent GLM.} The time-dependent GLM is expressed as $P(\Tdag_i \le t_0 \mid \bZ_i) = g\left(\alpha(t_0)+ \bbeta(t_0)\trans\bZ_i\right)$, where both $\alpha(t_0)$
 and $\bbeta(t_0)$ are functions of $t_0$. Given a $t_0$, these parameters can be estimated by the double IPW estimation.  Each observation is weighted by $\what_j\ast \omegahat_{t_0,i}$, where $\what_j$ is the weight  in equation (\ref{equ:what}) accounting for the missing values of $\bZ_i$ due to the sampling, and $\omegahat_{t_0,i}$ is the weight accounting for the missing disease status $I(\Tdag_i\le t_0)$ due to censoring. The censoring weight $\omegahat_{t_0,i}$ is given as $\omegahat_{t_0,i} =  \delta_i I(T_i \le t_0)/\Gschat(T_i) + I(T_i > t_0)/\Gschat(t_0)$, where $\Gschat(t)$ is a consistent estimator of $\Gsc(t)=P(C_i\geq t)$, the survival function of the censoring time. If the censoring is independent of both the event time and markers, $\Gschat(t)$ could be obtained by the Kaplan-Meier estimator \citep{kaplan1958nonparametric}. If the censoring depends on the markers $\bZ_i$, a PH model can be fit to estimate $P(C_i\geq t \mid \bZ_i)$.


\subsubsection{Accuracy parameters estimation}

The probability $\Pbbm_{t_0,i}$ in equation (\ref{equ:risk-model}) can be used as a risk score that classifies subjects into different risk categories. Given a cut-off value $c$, subjects with $\Pbbm_{t_0,i} \geq c$ are classified as the high-risk group, and the low-risk group consists of subjects with  $\Pbbm_{t_0,i} < c$. Several time-dependent accuracy measures have been proposed to evaluate the prediction performance of a risk score. In this paper, we consider the time-dependent TPR, FPR, PPV, and NPV. They are defined as: $\tpr_{t_0}(c) = Pr(\Pbbm_{t_0,i} > c \mid T^{\dag}_i \leq t_0)$, $\fpr_{t_0}(c) = Pr(\Pbbm_{t_0,i} > c \mid T^{\dag}_i > t_0)$, $\ppv_{t_0}(c) =Pr(T^{\dag}_i \leq t_0 \mid \Pbbm_{t_0,i} > c)$, and $\npv_{t_0}(c) = Pr(T^{\dag}_i > t_0 \mid \Pbbm_{t_0,i} \leq c)$. In addition, the time-dependent AUC is the area under the time-dependent ROC curve, which is a curve of $\tpr_{t_0}(c)$ versus $\fpr_{t_0}(c)$ over all possible values of $c$. The time-dependent AUC can be expressed as $\auc_{t_0}(c)=Pr(\Pbbm_{t_0,i} > \Pbbm_{t_0,j} \mid T^{\dag}_i \leq t_0, T^{\dag}_j \leq t_0)$, a conditional probability that, given a pair of an event and a non-event, the event has a higher risk score.

Let $\alphahat_{t_0}$ and $\bbetahat_{t_0}$ denote the IPW estimates of the model parameters under either the Cox PH model or the time-dependent GLM. With these estimates, we can calculate the estimated risk $\Pbbmhat_{t_0,i} = g(\alphahat_{t_0} + \bbetahat_{t_0} \trans\bZ_i)$.  The time-dependent accuracy measures described above can be estimated by the double IPW estimators: 
$$
\tprhat_{t_0}(c) = \frac{\sumiN \what_i \omegahat_{t_0,i} I(
\Pbbmhat_{t_0, i} > c)I(T_i  \le t_0)}{\sumiN\what_i \omegahat_{t_0,i} I(T_i  \le t_0)},\ \fprhat_{t_0}(c) = \frac{\sumiN \what_i \omegahat_{t_0,i} I(
\Pbbmhat_{t_0, i} > c)I(T_i  > t_0)}{\sumiN\what_i \omegahat_{t_0,i} I(T_i  > t_0)},
$$
$$\ppvhat_{t_0}(c) = \frac{\sumiN \what_i \omegahat_{t_0,i} I(\Pbbmhat_{t_0, i} > c)I(T_i  \le t_0)}{\sumiN\what_i \omegahat_{t_0,i} I(
\Pbbmhat_{t_0, i} > c)},\ \npvhat_{t_0}(c) = \frac{\sumiN \what_i \omegahat_{t_0,i} I(
\Pbbmhat_{t_0, i} \leq c)I(T_i  > t_0)}{\sumiN\what_i \omegahat_{t_0,i} I(
\Pbbmhat_{t_0, i} \leq c)},
$$ 
and
\begin{equation}\label{equ:auchat}
\auchat_{t_0}= \frac{\sumiN \sumjN \what_i \omegahat_{t_0,i}\what_j \omegahat_{t_0,j} I(
\Pbbmhat_{t_0, i} > \Pbbmhat_{t_0, j})I(T_i  \le t_0)I(T_j > t_0)}{\sumiN \sumjN \what_i \omegahat_{t_0,i} \what_j\omegahat_{t_0,j}I(T_i  \le t_0)I(T_j > t_0)}.
\end{equation}

\subsection{Asymptotic Variance of IPW Estimators}
In Appendix A, we show that for any $0<\pi_1\leq 1$, given the data, $E\left(\what_j \right) = 1$. As a result, the IPW estimators are consistent for the model parameters and accuracy parameters \citep{cai2012evaluating, cai2011nonparametric,zhou2015assessing}. They can be expressed as a weighted sum of independent zero-mean random variables that are functions of the data $\bD_j=(T_j,\delta_j,\bZ_j)$ (equation (\ref{equ:general-form}) of Appendix B.1). Thus, the IPW estimators are asymptotically normally distributed, and the asymptotic variance consists of two parts (equation (\ref{equ:V1}) - (\ref{equ:V3}) of Appendix B.1). One part accounts for the variability from (i) the randomness of $V_{1j}$'s and $V_{0j}$'s, (ii) the estimation of the censoring survival function, and  (iii) the estimation of the model and accuracy parameters. The other part accounts for between-case correlations of $V_{1j}$'s and between-control correlations of $V_{0j}$'s, which are not ignorable at the first order. The variance estimator given in \cite{samuelsen1997psudolikelihood} ignores the correlations and may lead to biased variance estimator.  

The expression of the asymptotic variance is complicated, and thus, a direct estimation is not feasible. Resampling procedures, such as bootstrap, cannot emulate the correlations, and thus, fail to estimate the variance accurately  \citep{gray2009weighted, cai2013resampling}. \cite{cai2013resampling} proposed a method that mimics the variances and correlations of $V_{0j}$'s via repeatedly perturbing these indicator variables. This method was designed for a standard NCC design where all the events are included as cases. However, when there is a sampling process for selecting cases, the perturbation of $V_{0j}$'s alone is not sufficient. Thus, we extend this procedure by perturbing both $V_{1j}$'s and $V_{0j}$'s to recover all the variances and correlations described above.

\subsection{Variance Estimation via Perturbation} \label{sec:perturbation}

Like bootstrap, the perturbation method creates a large number of perturbed counterparts for the estimator. We can calculate their empirical variance, which approximates the finite-sample variance of the estimator.  

For the IPW estimators of model parameters, their perturbed counterparts, denoted by $\alphahat_{t_0}^{\ast}$ and $\bbetahat_{t_0}^{\ast}$, are obtained by replacing the sampling and censoring weights, $\what_i$ and $\omegahat_{t_0,i}$ with their respective perturbed counterparts, $\whatast_i$ and $\omegahatast_{t_0,i}$. We will explain how to perturb these two weights in the next paragraph. For the IPW estimators of accuracy parameters, their counterparts are obtained using the weight  $\whatast_i*\omegahatast_{t_0,i}$ and the perturbed risk score  $\Pbbmhat^{\ast}_{t_0, i}=g\left(\alphahat_{t_0}^{\ast} + \bZ_i\trans{\bbetahat_{t_0}^{\ast}}\right)$. For example, the perturbed counterpart of $\auchat_{t_0}$ in equation (\ref{equ:auchat}) is given as 
$
\auchat_{t_0}^{\ast}= \frac{\sumiN \sumjN \whatast_i\omegahatast_{t_0,i}\whatast_j\omegahatast_{t_0,j} I\{
\Pbbmhat^{\ast}_{t_0, i} > \Pbbmhat^{\ast}_{t_0, j}, T_i  \le t_0, T_j > t_0)}{\sumiN \sumjN\whatast_i\omegahatast_{t_0,i} \whatast_j\omegahatast_{t_0,j}I(T_i  \le t_0, T_j > t_0)}
$.

The perturbed weights $\whatast_j$ and $\omegahatast_j$ are obtained as follows. Let $\{I_{ij},i=1,\cdots,N,j=1,\cdots,N\}$ be independent and identically distributed random variables with mean 1 and variance 1. The perturbed censoring weight $\omegahatast_{t_0,i}$ is given as $\omegahatast_{t_0,i}= \delta_i I(T_i \le t_0)/\Gschatast(T_i) + I(T_i > t_0)/\Gschatast(t_0)$, 
where $\Gschatast(t)$ is the estimate of $\Gsc(t)$ with each subject weighted by $I_{ii}$, $i=1,\cdots,N$.

The perturbed sampling weight $\what_j^{\ast}= \delta_{j}\frac{V_{1j}^{\ast}}{\pi_1^{*}} + (1-\delta_{j})\frac{V_{0j}^{*}}{\phat_{0j}^{*}}$, obtained by replacing the indicator variables $V_{1j}$ and $V_{0j}$ as well as their probabilities $\pi_1$ and $\phat_{0j}$ with their perturbed counterparts. Specifically, the perturbed counterpart of the case indicator is $V_{1j}^{\ast} = V_{1j}I_{jj}$. The selection probability $\pi_1$ can be written as $\pi_1 =  \sumiN \delta_iV_{1i}/\sumiN \delta_i$, and its perturbed counterpart is given as $\pi_1^{\ast}=\sumiN I_{ii}\delta_i V_{1i}/ \sumiN I_{ii}\delta_i$. Shown in Section 2.1, the control indicator $V_{0j}$ can be written as $V_{0j}=1-\prod_{i:j\in \Rscr_{i}}(1-V_{1i}V_{0j}^i)$, and its perturbed counterpart is give as $V_{0j}^{\ast}=1-\prod_{i:j\in \Rscr_{i}}(1-V_{1i}V_{0j}^iI_{ij})$. The probability $\phat_{0j}$ in equation (\ref{equ:p0j}) can be written as $\phat_{0j}=1-\prod_{i:j\in \Rsc_i} \left\{1-\frac{\sum_{l\in\Rsc_i}V_{0l}^i}{\nfrak_{i}-1}\delta_iV_{1i}\right\}$ where $m=\sum_{l\in\Rsc_i}V_{0l}^i$, and its perturbed counterpart is $\phat_{0j}^{*}=1-\prod_{i:j\in\Rscr_{i}} \left\{1-\frac{\sum_{l \in \Rscr_{i}}V_{0k}^{i}I_{il}}{\nfrak_{i}-1}\delta_{i}V_{1i}\right\}$. 

It is worth noting that our perturbation procedure is valid for all types of NCC designs, no matter whether cases consist of all the events or a subset of the events. When $\pi_1=1$, i.e., all the events are cases, $V_{1j}=\delta_j$, and $\pi_1^{\ast} = \pi_1=1$. The perturbed weights $\what_j^{\ast}$ become the same as those proposed in \cite{cai2013resampling} for standard NCC designs.

\section{Comparison with \cite{samuelsen1997psudolikelihood}'s Weight}  \label{sec:comparison}

As described earlier, Samuelsen's weight assigns the weight 1 to cases and assigns the weight of the inverse sampling probability to controls. It can be expressed as
\begin{equation}\label{equ:wbre}
\wbre_j = \delta_{j}V_{1j} + (1-\delta_{j}V_{1j})V_{0j}/\phat_{0j}.
\end{equation}
Table \ref{tab:weight-comparison} also lists $\wbre_j$ for the three groups of subjects in the sub-cohort: event cases, event controls, and non-event controls.

Consider using our new weight $\what_j$ and Samuelsen's weight $\wbre_j$ for the IPW estimation with the same NCC data. When all the events are selected as cases, i.e., $\pi_1=1$, $\what_j = \wbre_j$, and they lead to the same IPW estimators. However, when not all the events are cases, i.e., $\pi_1<1$, these two weights are different. Although their resulting IPW estimators are both consistent, they have different variations. 

Derived in equation (\ref{equ:general-variance}) of Appendix B.1, the asymptotic variance of IPW estimators is determined by the conditional variance of the weight given the data. As shown in Table \ref{tab:weight-comparison}, these two weights are the same for non-event controls. Thus, the difference between the two types of the weights lies in the conditional variances for event cases and event controls.  

For an event case, $\what_j=1/\pi_1$ but $\wbre_j=1$; $\what_j > \wbre_j$ (since $\pi_1<1$). This indicates that our method assigns heavier weights to event cases, compared to Samuelsen's weight. For an event control, $\what_j=0$ but $\wbre_j=1/\phat_{0j}$. Our method excludes event controls, but Samuelsen's weight generally assigns a large weight, because, explained in Remark 1, the probability $\phat_{0j}$ is small. 

As a result, for the events in the sub-cohort, the conditional variances of the two weights given the data are $Var(\wbre_j) =  (1-\pi_1) \E_{\bV_1}\left[\frac{1-\phat_{0j}}{\phat_{0j}} \mid V_{1j}=0\right]$ (equation (\ref{equ:appendix-varwbre-delta1})) and $Var(\what_j) = (1-\pi_1)/\pi_1$, where $\bV_1$ is the vector of $V_{1j}$'s for $j=1,\cdots,N$. Because the probability $\phat_{0j}$ is usually much smaller than $\pi_1$,  $Var(\wbre_i) > Var(\what_i)$. Thus, the IPW estimators using our new weight are more efficient, compared to Samuelsen's weight.

This conclusion is also confirmed by the numerical studies in Section 4. Furthermore, when $\pi_1$ gets smaller, the gap between the variances of these two IPW estimators becomes wider. It is because, with a smaller $\pi_1$, more events are not selected as cases. When an event with a short event time is selected as a control, the probability $\phat_{0j}$ is extremely small (Remark 1), and it inflates $Var(\wbre_i)$, and consequently, increase the variance of the IPW estimator.

\begin{remark}
As mentioned earlier, given the data, $E(\what_j)=1$. For Samuelsen's weight, for any $0<\pi_1\leq 1$, given the data, $E(\wbre_i \mid \bV_1)=\delta_j V_{1j} + (1-\delta_j V_{1j}) E[V_{0j}/\phat_{0j}]$. Since $E[V_{0j}/\phat_{0j}]=1$ \citep{samuelsen1997psudolikelihood}, $E(\wbre_i \mid \bV_1)=1$. Thus, given the data, $E(\wbre_i)=1$. When not all the events are cases, Samuelsen's weight can still lead to a consistent IPW estimator, but the estimator is less efficient than the one using our weight $\what_j$.

\end{remark}

\section{Numerical Studies}
We compare the two weights via numerical studies. Let $\theta$ denote a regression model coefficient parameter or an accuracy parameter. Let $\thetahat$ denote the IPW estimator using our new weight $\what_j$, and $\thetabre$ denote the estimator using Samelsen's weight $\wbre_j$.

\subsection{Simulation Study}\label{sec:simulation}

The purposes of the simulation study are two-fold. First, we compare the empirical bias and variance of $\thetahat$ versus $\thetabre$.  Second, we examine the validity of the proposed perturbation procedure for estimating the variance of the IPW estimator $\thetahat$. 


\subsubsection{Simulation setting}
We consider one clinical marker $Z_j$ that is measured on the full cohort, and one bio-marker $B_j$ that is measured only on the NCC sub-cohort. The marker $Z_j$ is first generated from the standard normal distribution $N(0,1)$, and the biomarker $B_j = Z_j + e_{\tiny B,j}$, where $e_{\tiny B,j} \sim N(0,1)$.  Given the two markers, the event time $T_j^{\dag}$ is obtained from $\log(T_j^{\dag}) = 1.5 - 0.25Z_j - 0.25 B_j + 0.5\epsilon_{\tiny T,j}$, where $\epsilon_{\tiny T,j}$ is generated from an extreme value distribution with the cumulative distribution function $F(x)=1-exp(-e^{x})$. The censoring time $C_j$ is generated from $C_j=\min\{0.1+C_{j1},C_{j2}\}$, where $C_{j1} \sim \mbox{Gamma}(2,2)$ and $C_{j2} \sim \mbox{Uniform}[0.5,2]$. Let $T_j=\min\{T_j^{\dag},C_j\}$ and $\delta_j=I(T_j^{\dag}\leq C_j)$. The censoring rate is about 97\%. 

The NCC design is often considered for large cohort such as the EPIC study \citep{jakszyn2006endogenous} which includes 521,457 subjects. Thus, we use $N=10,000$ as the size of the full cohort. To construct the sub-cohort sampling, we set $\pi_1$ (the percentage of events that are selected as cases) to be 20\%, 50\%, and 80\%. For each selected case, $m=3$ controls are selected from either (i) the case's risk set without matching, or (ii) the case's risk set with exact matching on a variable $M_1$ and matching up to $\pm 1$ on another variable $M_2$. The two matching variables $M_1$ and $M_2$ are generated as follows. Let $M_1=I(\tilde M_1>0.5)$ where $\tilde M_1 = \Phi(Z+e_1)$ with $e_1\sim N(0,1)$, and let $M_2$ be the closest integer to $\tilde M_2=5\Phi(B+e_2)$ where $e_2\sim N(0,1)$.

With each NCC data, we fit (i) the time-dependent GLM model with a logit link, and (ii) the Cox PH model to estimate the risk $P(T^{\dag}<1)$. Each model is evaluated on the following time-dependent accuracy measures: TPR, PPV, and NPV at a cutoff value making FPR=0.05, as well as AUC. Besides the two IPW estimators $\thetahat$ and $\thetabre$, we also obtain a ``reference" estimate using the full cohort data. Additionally, for the IPW estimator $\thetahat$, we implement the proposed perturbation method described in Section \ref{sec:perturbation} to estimate its standard error from $1000$ perturbed counterparts with $I_{ij}\sim Exp(1)$.

\subsubsection{Simulation results}
The results based on 1000 replications are shown in Tables~\ref{tab:Beta-Cox} (Cox model parameters), Table \ref{tab:ACC-Cox} (Cox accuracy parameters), Table \ref{tab:Beta-GLM} (GLM's model parameters), and Table \ref{tab:ACC-GLM} (GLM's accuracy parameters).  For each parameter,  we report its true value, the empirical bias of the full-cohort estimates, the empirical biases and empirical standard deviations (ESDs) of the estimates $\thetahat$ and $\thetabre$. 

\paragraph{$\thetahat$ versus $\thetabre$.} In general, under both models, the ESD of $\thetabre$ is twice to three times of that of $\thetahat$. This is consistent with the analytic comparisons of the estimator variances presented in Section \ref{sec:comparison}. Additionally, when $\pi_1$ gets smaller, the gap between these two ESDs gets wider, although they both increase because the sample size of the sub-cohort gets smaller. 

Under the Cox model, the biases of $\thetahat$ and $\thetabre$ are similar, and they are ignorable compared to their ESDs. However, under the time-dependent GLM, when $\pi_1=0.2$ and 0.5, the estimation using the weight $\wbre_j$ does not converge for a number of replications, and thus, the bias and ESD of $\thetabre$ go through the roof. 

These non-convergences are caused by the subjects who experience the event early and are selected as controls. In the GLM, their binary outcomes $I(T_i\leq t_0)$ are 1. On the other hand, as explained in Remark 1, their $\phat_{0j}$ are close to zero, leading to a huge weight $\wbre_j$. When this situation happens, the GLM using the weight $\wbre_j$ fails to converge. In comparison, the GLM using $\what_j$ still converges since $\what_j$ for those subjects are zero.

\paragraph{$\thetahat$ versus full-cohort estimates.} Compared to the full-cohort estimates, the bias of $\thetahat$ is larger when $\pi_1$ is small. However, for a larger $\pi_1$, such as 80\%, the biases of these two estimators are comparable.

\paragraph{Perturbation.} To confirm the validity of the proposed perturbation method, we compare the ESD with the average standard error (ASE) obtained from the perturbation. In addition, we report the empirical coverage probabilities of the 95\% confidence intervals using the perturbation standard error. The results show that ASE is close to the ESD, indicating that the proposed perturbation procedure is able to estimate the variance of the IPW estimator accurately. Therefore, the empirical coverage probabilities of the perturbation confidence intervals are close to the nominal level 95\%.

\subsection{Data Example: the Framingham Offspring Study}\label{sec:example}

To illustrate the proposed new weighting method, we sample sub-cohorts from the Framingham Heart Study \citep{Wawrzyniak2013}. This example uses the Offspring cohort, including 1501 males and 1644 females. On these participants, the time to a cardiovascular disease (CVD) event, the Framingham risk score (FRS), and a biomarker, C-reactive protein (CRP), are collected. 

The FRS was developed by \cite{wilson1998prediction} to estimate the 10-year CVD risk. This score is gender-specific, and based on several risk factors including age, systolic blood pressure, diastolic blood pressure, total cholesterol, high-density lipoprotein cholesterol, current smoking status, and diabetes status. The CRP is an inflammation biomarker, and shown to improve the prediction on top of the traditional risk variables from the FRS \citep{ridker2003clinical,cook2006effect}.

We sample two types of NCC sub-cohorts: (i) selecting $\pi_1=50$\% of the subjects who have developed CVD as cases and 1 control for each case, and (ii) selecting $\pi_1=25$\% of the subjects who have developed CVD as cases and 3 controls for each case. Their sample sizes are close: on average about 904 for sub-cohort (i) and 896 for sub-cohort (ii). For each type, we repeat the NCC sampling for 100 times.

We consider two prediction time $t_0 =$ 15 and 30 years with CVD event rate about 7\% and 34\% respectively. We obtain the estimates from both the Cox PH model and time-dependent GLM models with the gender-specific FRS and log transformed CRP as predictors. There is an ``outlier": for one sample with $\pi_1=25$\%, the estimation for the $15$-year GLM does not converge, due to the reasons we have explained above. We summarize the following results without this "outlier". 

\begin{figure}
\centering
\caption{Results of data example: boxplots of the estimates $\thetahat$ (IPW estimates with the new weight $\what_j$) and $\thetabre$ (IPW estimates with the Samuelsen's weight $\wbre_j$) for the marker effects under both the Cox PH model and time-dependent GLMs based on 100 NCC sub-cohorts.  The triangles inside the box represents the mean of the estimates and the dashed horizontal line represents the full-cohort estimates.}\label{fig:data_coef}
\vspace{0.2cm}
\includegraphics[width=1\textwidth]{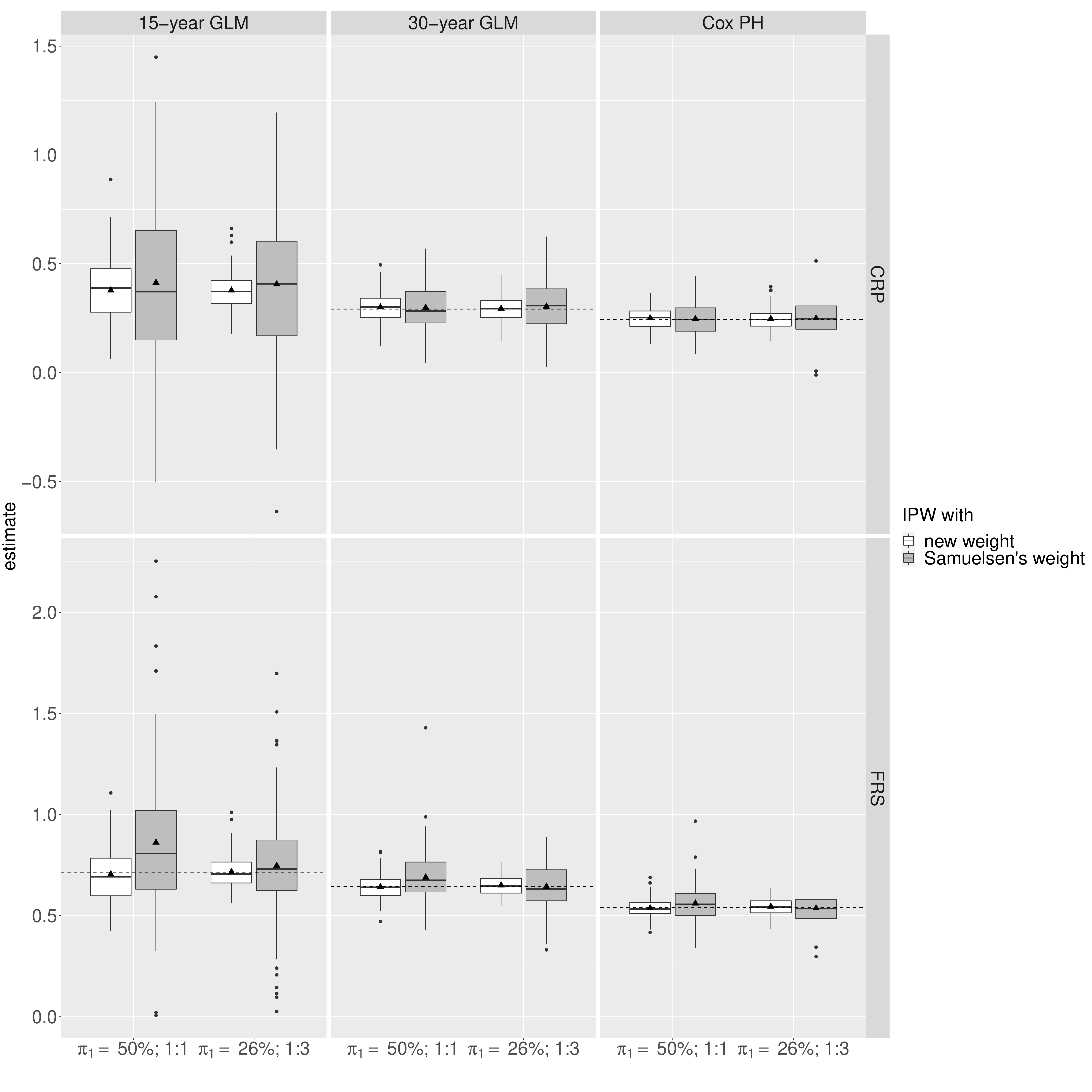}
\end{figure}

\begin{figure}
\centering
\caption{Results of data example: boxplots of the estimates $\thetahat$ (IPW estimates with the new weight $\what_j$) and $\thetabre$ (IPW estimates with the Samuelsen's weight $\wbre_j$) for the time-dependent accuracy parameters of the Cox PH model based on 100 NCC sub-cohorts. The triangles inside the box represents the mean of the estimates and the dashed horizontal line represents the full-cohort estimates.}\label{fig:data_Cox_acc}
\vspace{0.2cm}
\includegraphics[width=1\textwidth]{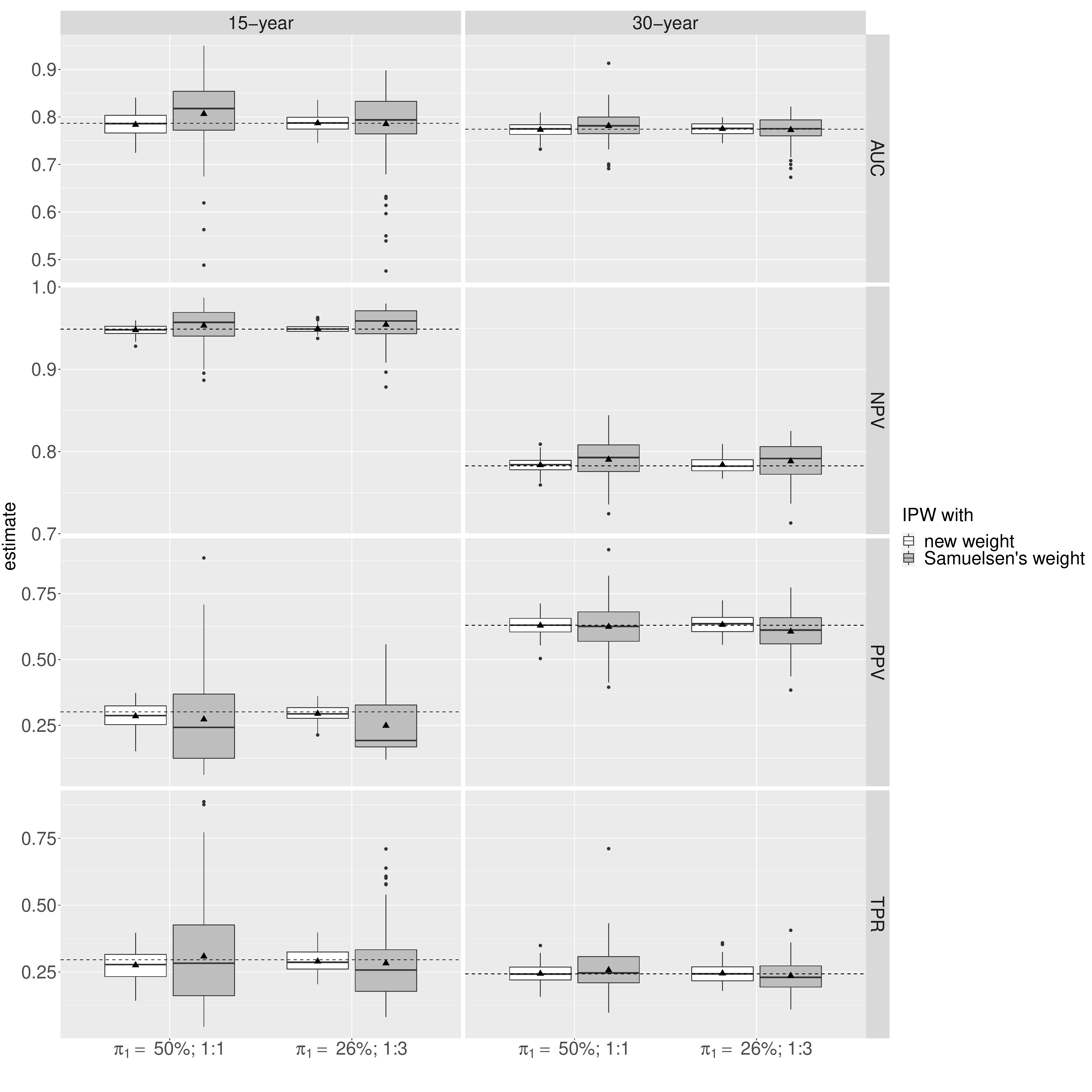}
\end{figure}

\begin{figure}
\centering
\caption{Results of data example: boxplots of the estimates $\thetahat$ (IPW estimates with the new weight $\what_j$) and $\thetabre$ (IPW estimates with the Samuelsen's weight $\wbre_j$) for the time-dependent accuracy parameters of the time-dependent GLM based on 100 NCC sub-cohorts. The triangles inside the box represents the mean of the estimates and the dashed horizontal line represents the full-cohort estimates.}\label{fig:data_GLM_acc}
\vspace{0.2cm}
\includegraphics[width=1\textwidth]{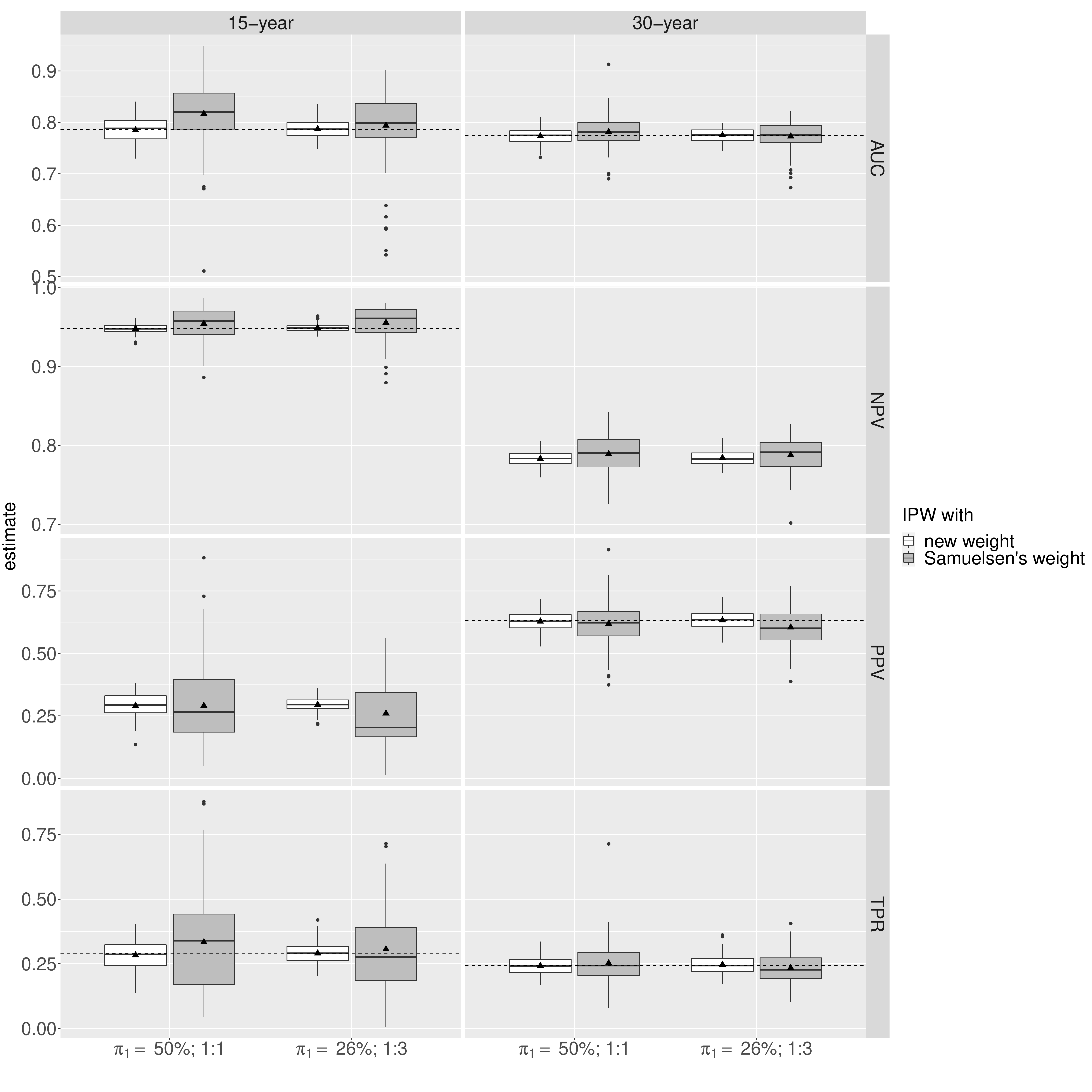}
\end{figure}

Figures \ref{fig:data_coef}, \ref{fig:data_Cox_acc}, and \ref{fig:data_GLM_acc} show the boxplots of the estimates $\thetahat$ and $\thetabre$. In each plot, the horizontal reference lines are the full-cohort estimates. For 30-year CVD outcome, the average of the estimates $\thetahat$ and $\thetabre$ are close, and both are similar to the full-cohort estimates. For 15-year CVD outcome with a limited number of events, $\thetabre$ can be biased as large as 20\% of the full-cohort estimates (the regression coefficient estimate for FRS under the 15-year GLM). As expected, the variation in $\thetabre$ is much larger than that of $\thetahat$. The difference of the variance between $\thetahat$ and $\thetabre$ is more significant for the marker effects  under the time-dependent GLM, compared to the Cox PH model. In addition, given that the sample sizes of the two types of sub-cohorts are similar, the variance of both $\thetahat$ and $\thetabre$ is mainly determined by $\pi_1$: the variance is smaller when $\pi_1$ is larger. 

\section{Discussion}\label{sec:discussion}

In this paper, we focus on a variation of NCC design in which cases consist of a subset, \textit{not all}, of the events. Such a design, although not generally considered in the statistical literature, is often encountered in practice for various reasons. In particular for biomarker studies, samples from cases may be more easily depleted and require careful preservation, and often, biospecimen collected on cases may not be sufficient or in good quality for processing. Providing valid estimating and inference procedures is urgently needed in the area. For such studies, we propose a new weight $\what_j$ in equation (\ref{equ:what}) for the IPW estimators of the model and accuracy parameters. Valid inference for such a study design is particularly challenging and to date has not been addressed. We propose a perturbation method to approximate the variances and between-subject correlations of the sampling indicator variables for both cases and controls. The simulation studies show that the standard error obtained from this method is close to the empirical standard deviation of the IPW estimator, and consequently, the coverage probabilities of the perturbation confidence intervals are close to the nominal level.

When all the events are cases, our new weight is equivalent to the Samuelsen weight $\wbre_j$ in equation (\ref{equ:wbre}), which was created for a standard NCC design. However, when not all the events are cases, $\what_j \neq \wbre_j$ for the events that are selected as cases or controls.  Both the analytical and numerical studies show that the IPW estimator with our new weight $\what_j$ is more efficient and robust under different risk models.  

Additionally, the proposed weighting method can be further extended to NCC studies where the cases are sampled via a more complex design, such as stratified sampling \citep{lu2018multi}. The weight $\what_j$ can be expressed as $\what_j =\delta_{j}V_{1j}/\phat_{1j} + (1-\delta_{j})V_{0j}/\phat_{0j}$, where $\phat_{1j}$ is the probability that subject $j$ is selected as a case based on the sampling scheme. Together, our proposed work could open the door for more efficient and practical biomarker studies.

\appendix
\numberwithin{equation}{subsection}
\section*{Appendices}
\renewcommand{\thesubsection}{\Alph{subsection}}

\subsection{Variance of Sampling Weights $\what_j$ and $\wbre_j$}

First, we derive the variance of the proposed new weights $\what_j = \delta_{j}V_{1j}/\pi_1 + (1-\delta_{j})V_{0j}/\phat_{0j}$. Let $\Dfrak$ include $T_i$, $\delta_i$, and the matching variable $\bM_i$ if the controls are selected from the matching risk set. Let $\bV_1=\{V_{11},\cdots,V_{1N}\}$. Since $\pi_1 = Pr(V_{1j}=1 \mid \Dfrak)$ and $\phat_{0j} = Pr(V_{0j}=1 \mid \Dfrak, \bV_1)$ , we can show that $\E\left(\what_j \mid \Dfrak,\bV_1\right) = \delta_{j}V_{1j}/\pi_1 + (1-\delta_{j})$, and consequently, $\E\left(\what_j \mid \Dfrak \right) = \delta_{j}+ (1-\delta_{j}) = 1$. Also, we can show that $\Var\left(\what_j \mid \Dfrak,\bV_1\right) =  (1-\delta_{j}) \frac{1-\phat_{0j}}{\phat_{0j}}$, and thus 
\begin{align*}
\Var\left(\what_j \mid \Dfrak\right) & = \Var_{\bV_1}\left[\E\left(\what_j \mid \Dfrak,\bV_1\right)\right] + \E_{\bV_1}\left[\Var\left(\what_j \mid \Dfrak,\bV_1\right)\right] \\
& = \delta_j \frac{1-\pi_1}{\pi_1} + (1-\delta_{j})\E_{\bV_1}\left(\frac{1-\phat_{0j}}{\phat_{0j}} \mid \Dfrak \right).
\end{align*}
For subjects with $\delta_j=0$, i.e., non-events, $\Var(\what_j) = \E_{\bV_1}\left(\frac{1-\phat_{0j}}{\phat_{0j}} \mid \Dfrak \right)$, accounting for the variability from sampling controls; for subjects with $\delta_j=1$, i.e., events, $\Var(\what_j) = \frac{1-\pi_1}{\pi_1}$, accounting for the variability from sampling cases. Note that even though events can be selected as controls, since $\what_j=0$, this process does not contribute to  $\Var(\what_j)$ for events. 

In addition, we derive the covariance of the weights from two different subjects,
\begin{align*}
\Cov\left(\what_i,\what_j \mid \Dfrak\right) & = \Cov_{\bV_1}\left[\E\left(\what_i \mid \Dfrak,\bV_1\right), \E\left(\what_j \mid \Dfrak,\bV_1\right)\right] + \E_{\bV_1}\left[\Cov\left(\what_i,\what_j \mid \Dfrak,\bV_1\right)\right] \\
& = \delta_i \delta_j \frac{Cov(V_{1i},V_{1j})}{\pi_1^2}+ (1-\delta_{i})(1-\delta_j)\E_{\bV_1}\left(\rhohat_{ij} \frac{1-\phat_{0i}}{\phat_{0i}}\frac{1-\phat_{0j}}{\phat_{0j}} \mid \Dfrak \right),
\end{align*}
where $\rhohat_{ij}$ is the conditional correlation between $V_{0i}$ and $V_{0j}$ given the data and $\bV_1$. The covariance accounts for the between-event correlations induced by sampling cases and the between-non-event correlations induced by sampling controls. The covariance between an event and a non-event is 0.

Next, we derive the variance of the Samuelsen's weight $\wbre_j =  \delta_{j}V_{1j} + (1-\delta_{j}V_{1j})V_{0j}/\phat_{0j}$. We have $\E\left(\wbre_j \mid \Dfrak,\bV_1 \right)=\delta_{j}V_{1j} + (1-\delta_{j}V_{1j}) = 1$, and thus, $\E\left(\wbre_j \mid \Dfrak \right)=1$. Also, $\Var\left(\wbre_j \mid \Dfrak,\bV_1 \right)=(1-\delta_{j}V_{1j}) \frac{1-\phat_{0j}}{\phat_{0j}}$, and thus,
\begin{align*}
\Var\left(\wbre_j \mid \Dfrak\right) & = \Var_{\bV_1}\left[\E\left(\wbre_j \mid \Dfrak,\bV_1\right)\right] + \E_{\bV_1}\left[\Var\left(\wbre \mid \Dfrak,\bV_1\right)\right] \\
& = \E_{\bV_1}\left[(1-\delta_{j}V_{1j}) \frac{1-\phat_{0j}}{\phat_{0j}} \mid \Dfrak\right].
\end{align*}
For non-events, $\Var(\wbre_j) = \E_{\bV_1}\left(\frac{1-\phat_{0j}}{\phat_{0j}} \mid \Dfrak \right)$, same as $\Var(\what_j)$; however, for events, 
\begin{equation}\label{equ:appendix-varwbre-delta1}
\Var(\wbre_j) =  \E_{\bV_1}\left[(1-V_{1j}) \frac{1-\phat_{0j}}{\phat_{0j}} \mid \Dfrak\right] = (1-\pi_1) \E_{\bV_1}\left[\frac{1-\phat_{0j}}{\phat_{0j}} \mid \Dfrak,V_{1j}=0\right]
\end{equation}
accounting for only the variability from sampling controls given that the event was not selected as a case, but not the variability from sampling cases.

Because $\E\left(\wbre_j \mid \Dfrak,\bV_1 \right) =1$, the covariance of the weights from two different subjects,
\begin{align*}
\Cov\left(\wbre_i,\wbre_j \mid \Dfrak\right) & =  \E_{\bV_1}\left[\Cov\left(\wbre_i,\wbre_j \mid \Dfrak,\bV_1\right)\right] \\
& = \E_{\bV_1}\left[ (1-\delta_{i}V_{1i})(1-\delta_jV_{1j})\rhohat_{ij} \frac{1-\phat_{0i}}{\phat_{0i}}\frac{1-\phat_{0j}}{\phat_{0j}} \mid \Dfrak \right].
\end{align*}
The between-non-event covariance of the Samuelsen's weights is same as the covariance of the new weights. However, the between-event covariance accounts for the correlation induced by sampling controls given neither are selected as cases. Additionally, the covariance between an event and a non-event is not zero for the Samuelsen's weight.

\subsection{Variance estimation for IPW estimators via perturbation}

Let $\theta$ denote a model parameter or an accuracy parameter described in this manuscript, and let $\thetahat$ denote its IPW estimator using our new weight. For simplicity, we suppress $t_0$ from the parameters, such as $\bgamma_{t_0}$ and $\auc_{t_0}$. Let $\overline \theta$ denote the limiting value of $\thetahat$, as $N\rightarrow \infty$.  Let $\Wschat_{\stheta} = \sqrt{N} (\thetahat-\overline \theta)$.   Let $\thetahat^{\ast}$ be the perturb counterpart of the estimate $\thetahat$, and let $\Wschat^{\ast}_{\stheta} = \sqrt{N} (\thetahat^{\ast}-\overline \theta)$.



\subsubsection{Asymptotic variance of $\Wschat_{\stheta}$}
 
 Following the supplementary material of \cite{zhou2015assessing}, we can show that the IPW estimates $\Wschat_{\stheta}$ under the Cox PH model or the time-dependent GLM model can be expressed in the following general form
\begin{equation}\label{equ:general-form}
\Wschat_{\stheta} \simeq N^{-1/2} \sumjN \bCsc_1(\bD_j) +  N^{-1/2} \sumjN (\what_j-1) \bCsc_2(\bD_j),
\end{equation}
where $a \simeq b$ denotes $a=b+o_p(1)$, and $\bCsc_1(\bD_j)$ and $\bCsc_2(\bD_j)$ are function of the data $\bD_j=(T_j,\delta_j,\bZ_j)$ with mean 0. The expressions of $\bCsc_1(\bD_j)$ and $\bCsc_2(\bD_j)$ under the Cox PH model and time-dependent GLM are given in Appendices \ref{appendix:cox} and \ref{appendix:glm}.

Based on this expression, the asymptotic variance of  $\Wschat_{\stheta}$ is given as 
\begin{equation}\label{equ:general-variance}
E\left[\bCsc_1(\bD_j)^{\bigotimes 2}\right] + E\left[\bCsc_2(\bD_j)^{\bigotimes 2} Var(\what_j\mid \Dfrak)\right] + E\left[\bCsc_2(\bD_i)\bCsc_2(\bD_j) Cov(\what_i,\what_j\mid \Dfrak)\right]. 
\end{equation}
This indicates that the variance of the IPW estimator is determined by the conditional variance and between-subject covariance of the weight given the data.

Using the results derived in Appendix A, the asymptotic variance can be expressed as $Var\left[\Wschat_{\stheta} \right] = \Vsc_1 + \Vsc_2 + \Vsc_3$, where
\begin{align}
\label{equ:V1}\Vsc_1 = & E\left[\bCsc_1(\bD_j)^{\bigotimes 2}\right], \\
\label{equ:V2} \Vsc_2 = & E\left[\bCsc_2(\bD_j)^{\bigotimes 2}\delta_j \frac{1-\pi_1}{\pi_1}\right]  + E\left[\bCsc_2(\bD_i)\bCsc_2(\bD_j)\delta_i\delta_j \frac{Cov(V_{1i},V_{1j}\mid \Dfrak)}{\pi_1^2}\right], \\
\nonumber \Vsc_3 = & E\left[\bCsc_2(\bD_j)^{\bigotimes 2}(1-\delta_j)E_{\bV_1}\left(\frac{1-\phat_{0j}}{\phat_{0j}} \mid \Dfrak\right)\right]  + \\
\label{equ:V3} &\quad \quad  E\left[\bCsc_2(\bD_i)\bCsc_2(\bD_j)(1-\delta_i)(1-\delta_j)E_{\bV_{1}}\left(\rhohat_{ij}\frac{1-\phat_{0i}}{\phat_{0i}}\frac{1-\phat_{0j}}{\phat_{0j}}\mid \Dfrak\right)\right].
\end{align}
Note that $\Vsc_2$ accounts for the variance and covariance induced by sampling cases, and $\Vsc_3$ accounts for the variance and covariance induced by sampling controls.

We next show that given the observed data  $\mathcal F =  \{T_j,\delta_j,V_j\bZ_j, V_{1j}, V_{0j}^i, i=1,\cdots,n, j=1,\cdots,n\}$,  $Var\left[\Wschat^{\ast}_{\stheta}-\Wschat_{\stheta}\right]$ converges to $Var\left[\Wschat_{\stheta}\right]$.

\subsubsection{Convergence of $Var\left[\Wschat^{\ast}_{\stheta}-\Wschat_{\stheta}\right]$ to $Var\left[\Wschat_{\stheta}\right]$}  

The $\Wschat^{\ast}_{\stheta}-\Wschat_{\stheta}$ can be expressed as
\begin{equation} \label{equ:ptb-form}
\Wschat^{\ast}_{\stheta}-\Wschat_{\stheta} \simeq N^{-1/2} \sumjN (I_{jj}-1) \bCsc_1(\bD_j) + N^{-1/2} \sumjN \left[(\what^{\ast}_j-\what_j) - (I_{jj}-1)\right]\bCsc_2(\bD_j).
\end{equation}
It can be further re-written as $\Wschat^{\ast}_{\stheta}-\Wschat_{\stheta} \simeq W_1 + W_2 + W_3$, where 
\begin{align*}
W_1 &= N^{-1/2} \sumjN (I_{jj}-1) \bCsc_1(\bD_j),\\
W_2 & =N^{-1/2} \sumjN \delta_j(I_{jj}-1)\left(\frac{V_{1j}}{\pi_1} - 1\right) \bCsc_2(\bD_j) +  N^{-1/2} \sumjN  \delta_{j} \left( \frac{V_{1j}I_{jj}}{\pi_1^{*}} -\frac{V_{1j}I_{jj}}{\pi_1}\right) \bCsc_2(\bD_j),\\
W_3 &= N^{-1/2} \sumjN 
(1-\delta_{j})\left[\left(\frac{V_{0j}^{*}}{\phat_{0j}} -\frac{V_{0j}}{\phat_{0j}}\right)-(I_{jj}-1)\right]  \bR_{\sgamma}(\bD_j) \\
& \hskip 7cm + N^{-1/2} \sumjN (1-\delta_{j})\left(\frac{V_{0j}^{*}}{\phat_{0j}} -\frac{V_{0j}}{\phat_{0j}}\right)  \bR_{\sgamma}(\bD_j)
\end{align*}

Conditional on the observed data $\mathcal F$, $\bCsc_1(\bD_j)$ in $W_1$ serves as constants but $I_{jj}$'s are the random variable. Thus, the conditional expectation and variance of $W_1$ are $E\left[W_1 \mid \mathcal F\right] = N^{-1/2} \sumjN \bCsc_1(\bD_j)E(I_{jj}-1)=0$ and 
$$
Var\left[W_1 \mid \mathcal F\right] = N^{-1}\sumjN \bCsc_1(\bD_j)^{\bigotimes 2} Var[I_{jj}-1] = N^{-1}\sumjN \bCsc_1(\bD_j)^{\bigotimes 2}, 
$$
which converges to $\Vsc_1$ in equation (\ref{equ:V1}). Using similar arguments in \cite{agniel2016estimation}, we can show that the variance of $W_2$ conditional on $\Ffrak$ converges to $\Vsc_2$ in equation (\ref{equ:V2}) by setting all the $\delta$ to be 1 in \cite{agniel2016estimation}. Following the arguments in \cite{cai2013resampling}, the variance of $W_3$ conditional on $\Ffrak$ converges to $\Vsc_3$ in equation (\ref{equ:V3}). Thus, the variance of $\Wschat_{\stheta}^{\ast} - \Wschat_{\stheta}$ conditional on $\Ffrak$, converges to the variance of $\Wschat_{\stheta}$.

Next, we will present the derivation of the two general forms (\ref{equ:general-form}) and (\ref{equ:ptb-form}) for the estimates and their corresponding perturbed counterparts of the model parameters and accuracy parameters under the Cox PH model and time-dependent GLM model.

\subsubsection{The Cox PH Model}\label{appendix:cox} \cite{zhou2015assessing} shows that the model parameter estimator $\Wschat_{\sgamma}$ of the Cox PH model can be expressed as 
\begin{equation*}
\Wschat_{\sgamma} \simeq N^{-1/2} \sumjN \what_j \bR_{\sgamma}(\bD_j) = N^{-1/2} \sumjN \bR_{\sgamma}(\bD_j) + N^{-1/2} \sumjN (\what_j-1) \bR_{\sgamma}(\bD_j),
\end{equation*}
where $\bR_{\sgamma}(\bD_j)$ is a function of $\bD_j$ with mean 0 and its expression is given in Section C of the supplementary material of \cite{zhou2015assessing}. In terms of the general form (\ref{equ:general-form}), $\bCsc_1(\bD_j)=\bCsc_2(\bD_j)=\bR_{\sgamma}(\bD_j)$. Similarly, the perturb counterpart can be expresses as $\Wschat^{\ast}_{\sgamma} \simeq N^{-1/2} \sumjN \what_j^{\ast} \bR_{\sgamma}(\bD_j)$, and 
\begin{align}
\nonumber \Wschat^{\ast}_{\sgamma} - \Wschat_{\sgamma} & \simeq N^{-1/2} \sumjN (\what_j^{\ast}-\what_j) \bR_{\sgamma}(\bD_j)\\
\label{equ:cox-beta-ptb} & =  N^{-1/2} \sumjN (I_{jj}-1) \bR_{\sgamma}(\bD_j) + N^{-1/2} \sumjN \left[(\what^{\ast}_j-\what_j) - (I_{jj}-1)\right]\bR_{\sgamma}(\bD_j). 
\end{align}

\cite{zhou2015assessing} also show that the double IPW estimator of the accuracy measure, such as time-dependent AUC, can be expressed as
\begin{align}
\nonumber \Wschat_{\sauc} & \simeq N^{-1/2} \sumjN \what_j \omegahat_{t_0,j} \bR_{\sauc}(\bD_j) + \mathbbm A \left[N^{-1/2} \sumjN \what_j  \bR_{\sgamma}(\bD_j)\right]\\
\label{equ:cox-auc} & \simeq N^{-1/2} \sumjN \what_j \left[\omega_{t_0,j} \bR_{\sauc}(\bD_j) + \mathbbm A \bR_{\sgamma}(\bD_j)\right] + N^{-1/2} \sumjN \what_j (\omegahat_{t_0,j} - \omega_{t_0,j}) \bR_{\sauc}(\bD_j), 
\end{align}
where $\bR_{\sauc}(\bD_j)$ is also a function of $\bD_j$ with $E[\bR_{\sauc}(\bD_j)]=0$, $\mathbbm A= \partial \auc_{t_0}/\partial \bgamma_{t_0}$, and $\omega_{t_0,j}=\delta_j I(T_j<t_0)/\Gsc(T_j) + I(T_j>t_0)/\Gsc(t_0)$. The expressions of $\bR_{\sauc}(\bD_j)$ and $\mathbbm A$ are given in Section B.2 of the supplementary material of \cite{zhou2015assessing}. \cite{zhou2015assessing} also show that 
\begin{equation}\label{equ:whatC}
\omegahat_{t_0,j} - \omega_{t_0,j} \simeq N^{-1}\sumiN \mathcal U^{C}_i(\bD_j).
\end{equation}
and consequently, $N^{-1/2} \sumjN \what_j (\omegahat_{t_0,j} - \omega_{t_0,j}) \bR_{\sauc}(\bD_j)\simeq N^{-1/2} \sumjN E\left[\mathcal U^{C}_j(\bD_i)\bR_{\sgamma}(\bD_i) \right]$. Thus, $\Wschat_{\sauc}$ can be expressed as the general form (\ref{equ:general-form}), where 
\begin{equation}\label{equ:cox-auc-Csc}
\bCsc_2(\bD_j) =\omega_{t_0,j} \bR_{\sauc}(\bD_j) + \mathbbm A \bR_{\sgamma}(\bD_j),\ \text{and}\, \bCsc_1(\bD_j) = \bCsc_2(\bD_j) + E\left[\mathcal U^{C}_j(\bD_i)\bR_{\sgamma}(\bD_i) \right]. 
\end{equation}

Similar to equation (\ref{equ:cox-auc}), the perturb counterpart can be expressed as
\begin{align}
\nonumber \Wschat^{\ast}_{\sauc} & \simeq N^{-1/2} \sumjN \what_j^{\ast} \omegahatast_{t_0,j}\bR_{\sauc}(\bD_j) + \mathbbm A \left[N^{-1/2} \sumjN \what_j^{\ast}  \bR_{\sgamma}(\bD_j)\right]\\
\nonumber & = N^{-1/2} \sumjN \what_j^{\ast} (\omegahatast_{t_0,j}-\omega_{t_0,j})\bR_{\sauc}(\bD_j) + N^{-1/2} \sumjN \what_j^{\ast} \left[ \omega_{t_0,j}\bR_{\sauc}(\bD_j) + \mathbbm A \bR_{\sgamma}(\bD_j)\right]\\
\label{equ:cox-auc-ptb}& \simeq N^{-1/2} \sumjN \what_j (\omegahatast_{t_0,j}-\omega_{t_0,j})\bR_{\sauc}(\bD_j) + N^{-1/2} \sumjN \what_j^{\ast} \left[ \omega_{t_0,j}\bR_{\sauc}(\bD_j) + \mathbbm A \bR_{\sgamma}(\bD_j)\right],
\end{align}
and
\begin{align*}
\Wschat^{\ast}_{\sauc} - \Wschat_{\sauc} & \simeq N^{-1/2} \sumjN (\what_j^{\ast} - \what_j)\left[ \omega_{t_0,j}\bR_{\sauc}(\bD_j) + \mathbbm A \bR_{\sgamma}(\bD_j)\right] \\
& \quad \quad + N^{-1/2} \sumjN \what_j (\omegahatast_{t_0,j}-\omegahat_{t_0,j})\bR_{\sauc}(\bD_j).
\end{align*}
Following equation (\ref{equ:whatC}), it can be shown that 
\begin{equation}\label{equ:whatC-ptb}
\omegahatast_{t_0,j}-\omegahat_{t_0,j} \simeq N^{-1}\sumiN (I_{ii}-1)\mathcal U^{C}_i(\bD_j). 
\end{equation}
Thus, $\Wschat^{\ast}_{\sauc} - \Wschat_{\sauc}$ can be expressed as
\begin{align*}
\Wschat^{\ast}_{\sauc} - \Wschat_{\sauc} & \simeq N^{-1/2} \sumjN (\what_j^{\ast} - \what_j)\left[ \omega_{t_0,j}\bR_{\sauc}(\bD_j) + \mathbbm A \bR_{\sgamma}(\bD_j)\right] \\
& \quad \quad + N^{-1/2} \sumjN (I_{jj}-1) E\left[\mathcal U^{C}_j(\bD_i)\bR_{\sauc}(\bD_i) \right].
\end{align*}
Also following the argument in equation (\ref{equ:cox-beta-ptb}), we can express $\Wschat^{\ast}_{\sauc} - \Wschat_{\sauc} $ in the general form (\ref{equ:ptb-form}) where $\bCsc_1(\bD_j)$ and $\bCsc_2(\bD_j)$ are given in equation (\ref{equ:cox-auc-Csc}).

\subsubsection{Time-dependent GLM} \label{appendix:glm} 

\cite{zhou2015assessing} show that the double IPW estimator $\Wschat_{\sgamma}$ can be expressed as $\Wschat_{\sgamma} \simeq N^{-1/2} \sumjN \what_j \omegahat_{t_0,j} \bR_{\sgamma}(\bD_j)$ and the expression of $\bR_{\sgamma}(\bD_j)$ is given in Section B.2 of the supplementary material of \cite{zhou2015assessing}. It can be rewritten as
$$
\Wschat_{\sgamma} \simeq N^{-1/2} \sumjN \what_j \omega_{t_0,j} \bR_{\sgamma}(\bD_j) + N^{-1/2} \sumjN (\omegahat_{t_0,j} - \omega_{t_0,j}) \what_j \bR_{\sgamma}(\bD_j),
$$
With the equation (\ref{equ:whatC}), we can express $\Wschat_{\sgamma}$ in the general form (\ref{equ:general-form}) in which $\bCsc_2(\bD_j) = \omega_{t_0,j} \bR_{\sgamma}(\bD_j)$ and $\bCsc_1(\bD_j)=\bCsc_2(\bD_j) + E\left[\mathcal U^{C}_j(\bD_i)\bR_{\sgamma}(\bD_i) \right]$. Similarly, the perturbed estimate can be expressed as $\Wschat_{\sgamma}^{\ast} \simeq N^{-1/2} \sumjN \what_j^{\ast} \omegahatast_{t_0,j} \bR_{\sgamma}(\bD_j)$. Following the derivation in (\ref{equ:cox-auc-ptb}), it can be expressed as $\Wschat_{\sgamma}^{\ast} \simeq N^{-1/2} \sumjN \what_j^{\ast} \omega_{t_0,j} \bR_{\sgamma}(\bD_j) + N^{-1/2} \sumjN \what_j (\omegahatast_{t_0,j}-\omega_{t_0,j})\bR_{\sgamma}(\bD_j)$, and $\Wschat_{\sgamma}^{\ast} - \Wschat_{\sgamma}$ can be expressed as 
$$
\Wschat_{\sgamma}^{\ast} - \Wschat_{\sgamma} \simeq 
N^{-1/2} \sumjN (\what_j^{\ast}-\what_j) \omega_{t_0,j} \bR_{\sgamma}(\bD_j) + N^{-1/2} \sumjN (\omegahatast_{t_0,j} - \omegahat_{t_0,j}) \what_j \bR_{\sgamma}(\bD_j).
$$
With the expression in equation (\ref{equ:whatC-ptb}), $\Wschat_{\sgamma}^{\ast} - \Wschat_{\sgamma}$ can be expressed as 
\begin{align*}
\Wschat_{\sgamma}^{\ast} - \Wschat_{\sgamma} & \simeq  N^{-1/2} \sumjN (I_{jj}-1) \left\{\omega_{t_0,j} \bR_{\sgamma}(\bD_j)  + E\left[\mathcal U^{C}_j(\bD_i)\bR_{\sgamma}(\bD_i) \right]\right\} \\
& \quad \quad + N^{-1/2} \sumjN \left[(\what_j^{\ast}-\what_j) - (I_{jj}-1)\right]\omega_{t_0,j} \bR_{\sgamma}(\bD_j). 
\end{align*}

\cite{zhou2015assessing} show that the double IPW estimator of the accuracy measure, such as AUC, can be expressed as $\Wschat_{\sauc} \simeq N^{-1/2} \sumjN \what_j \omegahat_{t_0,j} \left[\bR_{\sauc}(\bD_j) + \mathbbm{A} \bR_{\sgamma}(\bD_j)\right]$. Similar to the expression $\Wschat_{\sgamma}$, it can be expressed in the general form (\ref{equ:general-form}) in which $\bCsc_2(\bD_j) = \omega_{t_0,j} \left[\bR_{\sauc}(\bD_j) + \mathbbm{A} \bR_{\sgamma}(\bD_j)\right]$ and $\bCsc_1(\bD_j)=\bCsc_2(\bD_j) + E\left\{\mathcal U^{C}_j(\bD_i)\left[\bR_{\sauc}(\bD_i) + \mathbbm{A} \bR_{\sgamma}(\bD_i)\right]\right\}$. Also, the perturbed counterpart $\Wschat_{\sauc}^{\ast} - \Wschat_{\sauc}$ can be expressed in the form (\ref{equ:ptb-form}).

\bibliographystyle{apalike}
\bibliography{NCC_ref}

\newpage

\begin{table}[!h]
\centering
\caption{Weights $\what_j$ and $\wbre_j$ for three groups of subjects in the sub-cohort: (i) events that are selected as cases, referred to as event cases, (ii) events that are selected as controls, referred to as event controls, (iii) non-events that are selected as controls, referred to as non-event controls}\label{tab:weight-comparison}
\vspace{0.2cm}
\begin{tabular}{|r|c|c|c|}
\hline \hline
& $(\delta_j,V_{1j},V_{0j})$ & $\what_j$ & $\wbre_j$ \\
\hline
event cases & $(1,1,0)$ & $1/\pi_1$ & 1 \\
\hline
event controls & $(1,0,1)$ & 0 & $1/\phat_{0j}$ \\
\hline
non-event controls & $(0,0,1)$ & $1/\phat_{0j}$ & $1/\phat_{0j}$ \\
\hline\hline
\end{tabular}
\end{table}

 \begin{table}[!h]
  \centering 
  \caption{Estimates of $\beta_{Z}$ and $\beta_{B}$ under the Cox PH model. For each parameter, the results include its true value, bias of the full-cohort estimates, bias of $\thetabre$ with the empirical standard deviation (ESD, in parentheses), bias of $\thetahat$ with the ESD (in parentheses), and the average standard error of $\thetahat$ obtained from 1000 perturbed counterparts (pASE) with the empirical coverage probabilities (Cov.Prob., in parentheses) of the 95\% confidence interval (CI) using the perturbation standard error.}\label{tab:Beta-Cox}
  \vspace{0.2cm}
  \begin{tabular}{rcccccc}
\hline \hline
& True & Bias (full-cohort) & $\thetabre$: Bias (ESD) & $\thetahat$: Bias (ESD) & pASE (Cov.Prob.)\\
\multicolumn{6}{c}{}\\
 & \multicolumn{5}{c}{$\pi_1=0.2$ Without Matching}\\
$\beta_Z$ & 0.5 & -1.9e-03 & -6.3e-03 (0.390) & 8.1e-03 (0.154) & 0.149 (0.94) \\
$\beta_B$ & 0.5 &  3.0e-04 &  1.2e-02 (0.277) & 6.2e-03 (0.113) & 0.106 (0.93) \\
\multicolumn{6}{c}{}\\
 & \multicolumn{5}{c}{$\pi_1=0.2$ With Matching}\\
$\beta_Z$ & 0.5 & -5.3e-05 & 1.0e-02 (0.362) & 6.3e-03 (0.140) & 0.138 (0.94) \\
$\beta_B$ & 0.5 &  8.8e-04 & 3.4e-02 (0.269) & 1.4e-03 (0.101) & 0.098 (0.95) \\
\multicolumn{6}{c}{}\\
\hline\\
 & \multicolumn{5}{c}{$\pi_1=0.5$ Without Matching}\\
$\beta_Z$ & 0.5 & 3.5e-04 & 3.9e-03 (0.218) & 2.2e-03 (0.100) & 0.095 (0.94) \\
$\beta_B$ & 0.5 & 1.4e-03 & 7.3e-03 (0.157) & 8.0e-03 (0.073) & 0.068 (0.93) \\
\multicolumn{6}{c}{}\\
  & \multicolumn{5}{c}{$\pi_1=0.5$ With Matching}\\
$\beta_Z$ & 0.5 &  3.3e-03 & 4.2e-03 (0.221) & 3.1e-03 (0.090) & 0.087 (0.94) \\
$\beta_B$ & 0.5 & -9.4e-04 & 1.1e-02 (0.159) & 1.8e-03 (0.064) & 0.062 (0.94) \\
\multicolumn{6}{c}{}\\
\hline\\
 & \multicolumn{5}{c}{$\pi_1=0.8$ Without Matching}\\
$\beta_Z$ & 0.5 & -4.8e-04 & -6.4e-03 (0.125) & 6.9e-04 (0.079) & 0.075 (0.94) \\
$\beta_B$ & 0.5 &  1.2e-03 &  8.9e-03 (0.094) & 4.1e-03 (0.055) & 0.054 (0.95) \\
\multicolumn{6}{c}{}\\
   & \multicolumn{5}{c}{$\pi_1=0.8$ With Matching}\\
$\beta_Z$ & 0.5 & -1.4e-03 & -8.4e-03 (0.126) & -6.9e-04 (0.071) & 0.069 (0.95) \\
$\beta_B$ & 0.5 & -1.3e-04 &  8.3e-03 (0.088) &  1.3e-03 (0.051) & 0.049 (0.94) \\
\hline \hline
\end{tabular}
\end{table}

\newpage
\begin{table}[!h]
  \centering 
  \caption{Estimates of the time-dependent AUC as well as the time-dependent TPR, NPV and PPV at the cutoff value such that FPR=0.05 under the Cox PH model. For each parameter, the results include its true value, bias of the full-cohort estimates, bias of $\thetabre$ with the ESD (in parentheses), bias of $\thetahat$ with the ESD (in parentheses), and the perturbation ASE (pASE) with the empirical coverage probabilities  (in parentheses) of the 95\% perturbation CIs.}\label{tab:ACC-Cox}
  \vspace{0.2cm}
{\small  \begin{tabular}{rcccccc}
\hline \hline
& True & Bias (full-cohort) & $\thetabre$: Bias (ESD) & $\thetahat$: Bias (ESD) & pASE (Cov.Prob.)\\
\multicolumn{6}{c}{}\\
 & \multicolumn{5}{c}{$\pi_1=0.2$ Without Matching}\\
AUC & 0.79 &  5.4e-05 &  6.4e-03 (0.070) &  7.1e-04 (0.026) & 0.028 (0.96) \\
TPR & 0.31 & -7.0e-04 &  3.0e-03 (0.159) & -7.4e-03 (0.069) & 0.070 (0.95) \\
NPV & 0.94 & -1.7e-04 &  9.8e-04 (0.023) & -6.7e-04 (0.008) & 0.008 (0.94) \\
PPV & 0.36 &  1.4e-04 & -1.7e-02 (0.137) &  1.1e-03 (0.056) & 0.060 (0.95) \\
\multicolumn{6}{c}{}\\
 & \multicolumn{5}{c}{$\pi_1=0.2$ With Matching}\\
AUC & 0.79 &  1.3e-04 &  2.2e-02 (0.072) & -4.3e-04 (0.027) & 0.028 (0.95) \\
TPR & 0.31 & -2.4e-04 &  1.6e-02 (0.144) & -3.2e-03 (0.058) & 0.063 (0.95) \\
NPV & 0.94 &  1.9e-04 &  3.0e-03 (0.025) & -3.7e-04 (0.008) & 0.008 (0.96) \\
PPV & 0.36 & -9.1e-04 & -1.7e-02 (0.113) &  9.7e-04 (0.049) & 0.052 (0.96) \\
\multicolumn{6}{c}{}\\
\hline\\
 & \multicolumn{5}{c}{$\pi_1=0.5$ Without Matching}\\
AUC & 0.79 &  3.7e-04 &  3.9e-03 (0.037) &  6.0e-04 (0.018) & 0.018 (0.94) \\
TPR & 0.31 & -1.0e-03 &  7.4e-04 (0.091) & -3.3e-03 (0.044) & 0.044 (0.94) \\
NPV & 0.94 & -1.3e-04 & -6.0e-05 (0.014) & -3.5e-04 (0.005) & 0.005 (0.94) \\
PPV & 0.36 & -4.0e-04 & -3.4e-03 (0.081) &  4.3e-04 (0.035) & 0.036 (0.95) \\
\multicolumn{6}{c}{}\\
  & \multicolumn{5}{c}{$\pi_1=0.5$ With Matching}\\
AUC & 0.79 &  3.0e-04 & 5.6e-03 (0.043) &  7.6e-04 (0.018) & 0.018 (0.94) \\
TPR & 0.31 &  1.9e-04 & 7.3e-03 (0.076) & -3.2e-04 (0.039) & 0.040 (0.96) \\
NPV & 0.94 &  1.8e-04 & 4.6e-04 (0.014) & -2.2e-05 (0.005) & 0.005 (0.94) \\
PPV & 0.36 & -3.3e-04 & 3.2e-04 (0.061) &  8.6e-04 (0.032) & 0.033 (0.95) \\
\multicolumn{6}{c}{}\\
\hline\\
 & \multicolumn{5}{c}{$\pi_1=0.8$ Without Matching}\\
AUC & 0.79 &  2.5e-04 &  7.9e-04 (0.022) &  3.8e-04 (0.014) & 0.014 (0.96) \\
TPR & 0.31 & -1.3e-03 & -2.2e-03 (0.053) & -3.2e-03 (0.034) & 0.035 (0.95) \\
NPV & 0.94 & -1.7e-04 & -2.4e-04 (0.007) & -3.5e-04 (0.004) & 0.004 (0.96) \\
PPV & 0.36 & -5.8e-04 & -1.2e-03 (0.046) & -1.8e-04 (0.028) & 0.029 (0.95) \\
\multicolumn{6}{c}{}\\
   & \multicolumn{5}{c}{$\pi_1=0.8$ With Matching}\\
AUC & 0.79 & -3.3e-04 &  1.2e-03 (0.025) & -6.9e-04 (0.014) & 0.014 (0.95) \\
TPR & 0.31 & -9.8e-04 & -1.3e-03 (0.047) & -1.6e-03 (0.031) & 0.031 (0.96) \\
NPV & 0.94 & -2.4e-04 & -5.5e-04 (0.009) & -3.8e-04 (0.004) & 0.004 (0.95) \\
PPV & 0.36 &  1.4e-04 & -8.5e-04 (0.037) &  8.0e-04 (0.026) & 0.026 (0.95) \\
\hline \hline
\end{tabular}
}
\end{table}

\newpage
\begin{table}[!h]
  \centering 
  \caption{Estimates of the intercept, $\beta_{Z}$ and $\beta_{B}$ under the time-dependent GLM. For each parameter, the results include its true value, bias of the full-cohort estimates, bias of $\thetabre$ with the ESD (in parentheses), bias of $\thetahat$ with the ESD (in parentheses), and the perturbation ASE (pASE) with the empirical coverage probabilities  (in parentheses) of the 95\% perturbation CIs.}\label{tab:Beta-GLM}
  \vspace{0.2cm}
  \begin{tabular}{rcccccc}
\hline \hline
& True & Bias (full-cohort) & $\thetabre$: Bias (ESD) & $\thetahat$: Bias (ESD) & pASE (Cov.Prob.)\\
\multicolumn{6}{c}{}\\
 & \multicolumn{5}{c}{$\pi_1=0.2$ Without Matching}\\
Intercept & -3.0 & -1.4e-03 & -7.4e+12 (1.1e+14) & -1.9e-02 (0.160) & 0.164 (0.96) \\
$\beta_Z$ &  0.5 & -3.0e-03 & -2.3e+11 (5.4e+13) &  9.5e-03 (0.215) & 0.211 (0.95) \\
$\beta_B$ &  0.5 &  3.0e-03 &  4.5e+11 (4.0e+13) &  8.7e-03 (0.157) & 0.152 (0.95) \\
\multicolumn{6}{c}{}\\
 & \multicolumn{5}{c}{$\pi_1=0.2$ With Matching}\\
Intercept & -3.0 & -8.1e-03 & -3.4e+13 (2.2e+14) & -1.8e-02 (0.162) & 0.166 (0.96) \\
$\beta_Z$ &  0.5 & -6.2e-04 &  1.4e+11 (8.2e+13) &  1.1e-02 (0.205) & 0.199 (0.94) \\
$\beta_B$ &  0.5 &  2.6e-03 &  3.0e+12 (4.3e+13) &  3.0e-03 (0.143) & 0.143 (0.96) \\
\multicolumn{6}{c}{}\\
\hline\\
 & \multicolumn{5}{c}{$\pi_1=0.5$ Without Matching}\\
Intercept & -3.0 & -3.7e-03 & -1.3e+12 (4.1e+13) & -8.4e-03 (0.110) & 0.107 (0.95) \\
$\beta_Z$ &  0.5 & -2.8e-06 & -7.9e+11 (2.5e+13) &  1.5e-03 (0.143) & 0.133 (0.93) \\
$\beta_B$ &  0.5 &  2.1e-03 &  9.4e+11 (3.0e+13) &  8.2e-03 (0.100) & 0.096 (0.94) \\
\multicolumn{6}{c}{}\\
  & \multicolumn{5}{c}{$\pi_1=0.5$ With Matching}\\
Intercept & -3.0 & -8.2e-03 & -1.3e+12 (3.0e+13) & -1.4e-02 (0.109) & 0.108 (0.95) \\
$\beta_Z$ &  0.5 &  5.2e-03 & -1.8e+11 (5.2e+12) &  1.2e-02 (0.131) & 0.125 (0.93) \\
$\beta_B$ &  0.5 & -6.5e-04 &  1.3e+11 (3.5e+12) &  5.9e-04 (0.095) & 0.090 (0.94) \\
\multicolumn{6}{c}{}\\
\hline\\
 & \multicolumn{5}{c}{$\pi_1=0.8$ Without Matching}\\
Intercept & -3.0 & -2.1e-03 & -1.3e-02 (0.137) & -2.2e-03 (0.086) & 0.087 (0.96) \\
$\beta_Z$ &  0.5 & -1.5e-04 & -5.9e-03 (0.151) & -1.1e-03 (0.112) & 0.105 (0.93) \\
$\beta_B$ &  0.5 &  1.9e-03 &  1.0e-02 (0.117) &  4.8e-03 (0.081) & 0.076 (0.94) \\
\multicolumn{6}{c}{}\\
   & \multicolumn{5}{c}{$\pi_1=0.8$ With Matching}\\
Intercept & -3.0 &  1.5e-03 & -2.1e+12 (6.8e+13) &  2.9e-03 (0.087) & 0.088 (0.95) \\
$\beta_Z$ &  0.5 & -1.6e-03 & -1.5e+11 (4.8e+12) & -1.2e-03 (0.103) & 0.099 (0.94) \\
$\beta_B$ &  0.5 & -2.0e-04 &  3.2e+11 (1.0e+13) &  3.2e-05 (0.070) & 0.071 (0.96) \\
\hline \hline
\end{tabular}
\end{table}

\newpage
\begin{table}[!h]
  \centering 
  \caption{Estimates of the time-dependent AUC as well as the time-dependent TPR, NPV and PPV at the cutoff value such that FPR=0.05 under the time-dependent GLM. For each parameter, the results include its true value, bias of the full-cohort estimates, bias of $\thetabre$ with the ESD (in parentheses), bias of $\thetahat$ with the ESD (in parentheses), and the perturbation ASE with the empirical coverage probabilities (in parentheses) of the 95\% perturbation CIs.}\label{tab:ACC-GLM}
  \vspace{0.2cm}
{\small  \begin{tabular}{rcccccc}
\hline \hline
& True & Bias (full-cohort) & $\thetabre$: Bias (ESD) & $\thetahat$: Bias (ESD) & pASE (Cov.Prob.)\\
\multicolumn{6}{c}{}\\
 & \multicolumn{5}{c}{$\pi_1=0.2$ Without Matching}\\
AUC & 0.79 &  1.5e-04 &  8.0e-03 (0.074) &  1.4e-03 (0.026) & 0.027 (0.95) \\
TPR & 0.31 & -4.6e-04 &  6.0e-03 (0.160) & -5.4e-03 (0.070) & 0.071 (0.94) \\
NPV & 0.94 & -1.5e-04 &  1.2e-03 (0.023) & -5.1e-04 (0.008) & 0.008 (0.94) \\
PPV & 0.36 &  3.3e-04 & -1.5e-02 (0.138) &  2.4e-03 (0.056) & 0.060 (0.95) \\
\multicolumn{6}{c}{}\\
 & \multicolumn{5}{c}{$\pi_1=0.2$ With Matching}\\
AUC & 0.79 &  2.4e-04 &  2.1e-02 (0.084) &  3.4e-04 (0.027) & 0.028 (0.95) \\
TPR & 0.31 & -2.2e-04 &  1.5e-02 (0.148) & -1.8e-03 (0.058) & 0.064 (0.96) \\
NPV & 0.94 &  1.9e-04 &  2.8e-03 (0.025) & -2.6e-04 (0.008) & 0.008 (0.96) \\
PPV & 0.36 & -8.9e-04 & -2.0e-02 (0.115) &  1.9e-03 (0.049) & 0.053 (0.96) \\
\multicolumn{6}{c}{}\\
\hline\\
 & \multicolumn{5}{c}{$\pi_1=0.5$ Without Matching}\\
AUC & 0.79 &  4.8e-04 &  4.7e-03 (0.037) &  9.6e-04 (0.018) & 0.018 (0.94) \\
TPR & 0.31 & -7.2e-04 &  1.7e-03 (0.091) & -2.8e-03 (0.044) & 0.044 (0.94) \\
NPV & 0.94 & -1.0e-04 &  9.2e-06 (0.014) & -3.1e-04 (0.005) & 0.005 (0.94) \\
PPV & 0.36 & -1.8e-04 & -2.4e-03 (0.080) &  8.3e-04 (0.036) & 0.036 (0.95) \\
\multicolumn{6}{c}{}\\
  & \multicolumn{5}{c}{$\pi_1=0.5$ With Matching}\\
AUC & 0.79 &  4.1e-04 &  6.2e-03 (0.044) & 1.1e-03 (0.018) & 0.018 (0.94) \\
TPR & 0.31 &  4.1e-04 &  5.8e-03 (0.077) & 6.2e-05 (0.039) & 0.040 (0.96) \\
NPV & 0.94 &  2.0e-04 &  2.9e-04 (0.014) & 9.1e-06 (0.005) & 0.005 (0.93) \\
PPV & 0.36 & -1.8e-04 & -8.2e-04 (0.060) & 1.1e-03 (0.032) & 0.033 (0.95) \\
\multicolumn{6}{c}{}\\
\hline\\
 & \multicolumn{5}{c}{$\pi_1=0.8$ Without Matching}\\
AUC & 0.79 &  3.5e-04 &  1.1e-03 (0.022) &  5.6e-04 (0.014) & 0.014 (0.95) \\
TPR & 0.31 & -1.1e-03 & -1.5e-03 (0.053) & -2.3e-03 (0.034) & 0.035 (0.95) \\
NPV & 0.94 & -1.5e-04 & -1.9e-04 (0.007) & -2.8e-04 (0.004) & 0.004 (0.95) \\
PPV & 0.36 & -4.2e-04 & -7.5e-04 (0.047) &  3.4e-04 (0.028) & 0.029 (0.95) \\
\multicolumn{6}{c}{}\\
   & \multicolumn{5}{c}{$\pi_1=0.8$ With Matching}\\
AUC & 0.79 & -2.3e-04 &  1.5e-03 (0.026) & -4.9e-04 (0.014) & 0.014 (0.95) \\
TPR & 0.31 & -7.5e-04 & -5.6e-04 (0.047) & -1.0e-03 (0.031) & 0.032 (0.95) \\
NPV & 0.94 & -2.2e-04 & -4.8e-04 (0.009) & -3.4e-04 (0.004) & 0.004 (0.95) \\
PPV & 0.36 &  3.1e-04 & -1.4e-04 (0.037) &  1.2e-03 (0.026) & 0.026 (0.95) \\
\hline \hline
\end{tabular}
}
\end{table}

\end{document}